\documentclass{appolb}
\usepackage{epsfig}
\usepackage{amsmath}
\newcommand{\be}{\begin{equation}}
\newcommand{\ee}{\end{equation}}
\newcommand{\bea}{\begin{eqnarray}}
\newcommand{\eea}{\end{eqnarray}}

\newcommand{\ev}{ {\rm eV} } 
\newcommand{\kev}{ {\rm keV} } 
\newcommand{\mev}{ {\rm MeV} }

\newcommand{\tonethr}{\mbox{$\theta_{13}$}}
\newcommand{\ttwothr}{\mbox{$\theta_{23}$}}

\newcommand{\oscfreq}{\mbox{$\frac{\Delta m_{23}^2 L}{4E}$}}

\newcommand{\numu}{\mbox{$\nu_{\mu}$}}
\newcommand{\numubar}{\mbox{$\overline{\nu}_{\mu}$}}
\newcommand{\nutau}{\mbox{$\nu_{\tau}$}}

\newcommand{\nue}{\mbox{$\nu_{e}$}}
\newcommand{\nuebar}{\mbox{$\overline{\nu}_{e}$}}

\begin{document}
%\date{\today}
\pagestyle{plain}
%% uncomment the following line to get equations numbered by (sec.num)
%\eqsec
\newcount\eLiNe\eLiNe=\inputlineno\advance\eLiNe by -1
\title{Neutrino Oscillation Experiments at Fermilab%
\thanks{Presented at the Cracow Epiphany Conference on Neutrinos 
in Physics and Astrophysics, Cracow, January 6-9,2000}
}
\author{Adam Para
\address{Fermilab, Pine St., Batavia IL 60510, USA }}
\maketitle

\begin{abstract}
\ Neutrino oscillations provide an unique opportunity to probe physics
beyond the Standard Model. Fermilab is constructing two new neutrino
beams to provide a decisive test of two of the recent positive indications
for neutrino oscillations: MiniBOONE experiment will settle the LSND
controversy, MINOS will provide detailed studies of the region indicated
by the SuperK results. 
\end{abstract}

\section{Introduction}

Neutrinos are special. They are the only elementary fermions which are neutral, thus they can be their own antiparticles.
Their  masses (if non-zero) are several orders of magnitude  smaller than those of charged leptons 
or quarks. These two
facts may be, in fact, related: with both Dirac and Majorana, masses present,
the neutrino mass eigenstates will naturally split, with the light mass eigenstate being
\begin{equation}
m_{\nu} \sim \frac{m_{D}^{2}}{m_{R}} 
\label{mass}
\end{equation}

If the Dirac mass $m_D$ is of the order of a typical quark or charged lepton mass and the 
right-handed Majorana mass is of the order of the GUT scale, the left-handed neutrinos
would then have a mass well below $1$ \ev. 

Direct measurements of the neutrino 
masses are very difficult; experiments so far have been able to yield only upper limits.
 The best  limit can be set for the $\nuebar$ from tritium $\beta$ decay, which is of the order of $1$ \ev \cite{Bonn}. A very similar 
limit, although dependent on the assumed Majorana nature of the neutrino, 
is derived from the measured rates of the neutrino-less double beta decays.
Mass limits for other neutrino species are considerably worse: $m_{\numu}~ < 170$~\kev from the decay of charged pions and $m_{\nutau}~< 24$~\mev from the decay $\tau ~\rightarrow~5\pi^{\pm}+\nutau$. Neutrino oscillations offer the
only practical means to unravel details of the neutrino mass spectrum.

\section{Neutrino Oscillations}

If neutrinos have masses, we may expect in analogy with the quark sector, that
the weak interaction eigenstates are mixtures of the mass eigenstates $\nu_1,\nu_2,\nu_3$, with the CKM-like mixing matrix U \cite{maki} 

\begin{equation}
U=\left(
\begin{array}{ccc}
U_{e1} & U_{e2} & U_{e3} \\
U_{\mu1} & U_{\mu2} & U_{\mu3} \\
U_{\tau1} & U_{\tau2} & U_{\tau3}
\end{array}
\right)
\end{equation}
usually parametrized as 
\begin{equation}
 \left(
\begin{array}{ccc}
c_{13}c_{12} & c_{13}s_{12} & s_{13}e^{-i\delta} \\ 
-c_{23}s_{12}-s_{13}s_{23}c_{12}e^{i\delta} & c_{23}c_{12}-s_{13}s_{23}s_{12}e^{i\delta } & 
c_{13}s_{23} \\ 
s_{23}s_{12}-s_{13}c_{23}c_{12}e^{i\delta} & -s_{23}c_{12}-s_{13}c_{23}s_{12}e^{i\delta } & 
c_{13}c_{23}
\end{array}
\right)
\label{CKM}
\end{equation}
where $s_{ij}=\sin \vartheta _{ij}$ and $c_{ij}=\cos \vartheta _{ij}$ and $\vartheta _{ij}$ is the mixing angle of $\nu_i$ and $\nu_j$.

Differences of mass eigenvalues will lead, through differences in time evolution
of the components of the wave function, to the phenomenon of neutrino 
oscillations:\cite{pontecorvo}  the beam of neutrinos of a given flavor, say \numu, will
be observed as a mixture of all three neutrino flavors after a certain distance
$\rm L$. Frequency of these oscillation is governed by $\Delta m^2_{ij}=m_i^2-m_j^2$.
    
 For example, starting with a pure $\numu$ beam and assuming $|\Delta m_{12}^2| \ll |\Delta m_{
23}^2|$, we would expect probabilities of detecting of $\numu$, $\nue$ and $\nutau$ to be
\begin{eqnarray}
P\left( \nu_\mu \rightarrow  \nu_\mu \right) & = &
1 - 4U^2_{\mu3}\left( 1-U^2_{\mu3} \right) \sin^2 \left ( \oscfreq \right )  \cr
P\left(  \nu_\mu \rightarrow \nu_e \right) & = &  
 \sin^2 \ttwothr \, \sin^2 2 \tonethr \, \sin^2 \left ( \oscfreq \right ) \cr
P\left(  \nu_\mu \rightarrow \nu_\tau \right) & = &  
\cos^4 \tonethr \, \sin^2 2 \ttwothr \, \sin^2 \left ( \oscfreq \right )  
\label{oscprobs}
\end{eqnarray}
where $E$ is the neutrino beam energy and $L$ is the distance of the detector from the neutrino source.

 At present, there are three experimental indications, shown in Fig.\ref{fig:three_osc},  that the neutrino oscillations might, in fact, occur in nature:
\begin{figure}[h]
\begin{center}
\epsfig{file=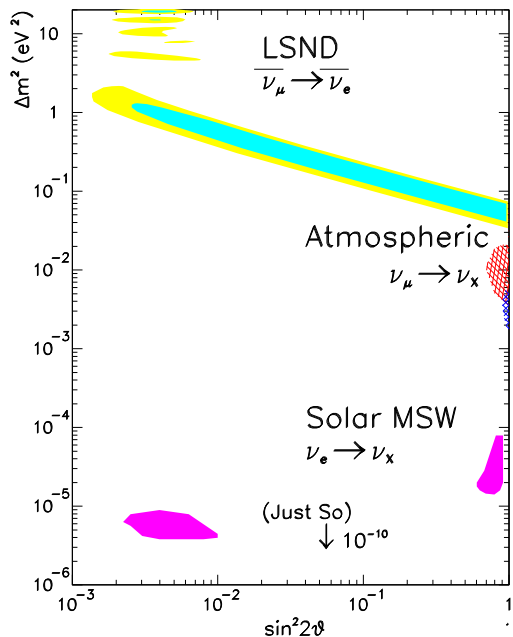,height=9cm}
\caption{Summary of possible indications for neutrino oscillations}\label{fig:three_osc}
\end{center}
\end{figure}

\begin{enumerate}
\item Solar neutrino deficit. 

The flux  of solar $\nue$ measured in several 
experiments is about 50\% of the flux expected in the Standard Solar 
Model. This large discrepancy is unlikely to be caused by our ignorance of the
physics of the Sun; it can be interpreted as a result of $\nue \rightarrow \nu_x$  oscillations. The $\Delta m^2 $ responsible for these oscillations would be of the order of $10^{-10}~\rm{eV}^2$ if the oscillations occur in a vacuum,
or $10^{-5}-10^{-4}~\rm{eV}^2$ if the oscillation occur in matter via the MSW effect.   
\item Atmospheric neutrinos.  

The SuperKamiokande detector\cite{SuperK} has been used to detect interactions of atmospheric neutrinos. The results show
depletion of the $\nu _{\mu }$ interaction rate as a function of the zenith
angle, while the $\nu _{e}$ interaction rate is consistent with the
expectations\cite{SuperK}, as shown in Fig.\ref{fig:superk}.  The observed
depletion is consistent with the hypothesis of neutrino oscillations and  
 strongly suggests $\nu _{\mu }\rightarrow \nu _{\tau }$
oscillations with a very large mixing angle and a $\Delta m^{2}$ in the
range $0.003-0.01eV^{2}$ \cite{kielcz}    
\begin{figure}[h]
\begin{center}
\epsfig{file=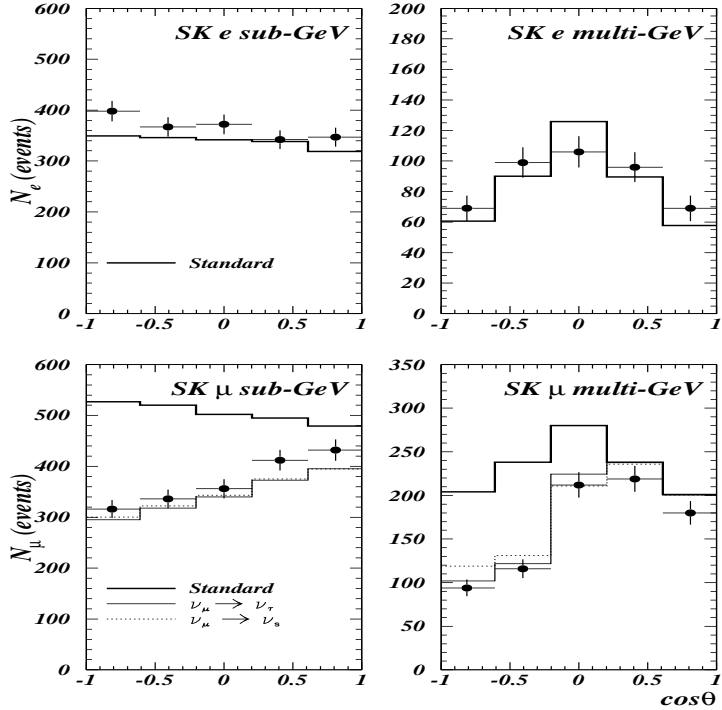,height=9.5cm,width=10cm}
\caption{Zenith angle distribution of atmospheric $\nue$ and $\numu$ events
observed in SuperKamiokande experiment}\label{fig:superk}
\end{center}
\end{figure}
\item LSND Experiment. 

An 800 MeV proton beam at LAMPF was used to produce 
pions, which were subsequently stopped in the absorber. A liquid scintillator detector 
recorded an excess of $82.8\pm23.7$ $\nuebar$ interactions above the expected 
background of $17.3\pm4$ events. These interactions are consistent with the 
hypothesis of $\numubar 
\rightarrow \nuebar$ oscillations (Fig.\ref{fig:lsnd}), if the oscillation 
parameters are in the region shown in Fig.\ref{fig:three_osc}. 
\begin{figure}[h]
\begin{center}
\epsfig{file=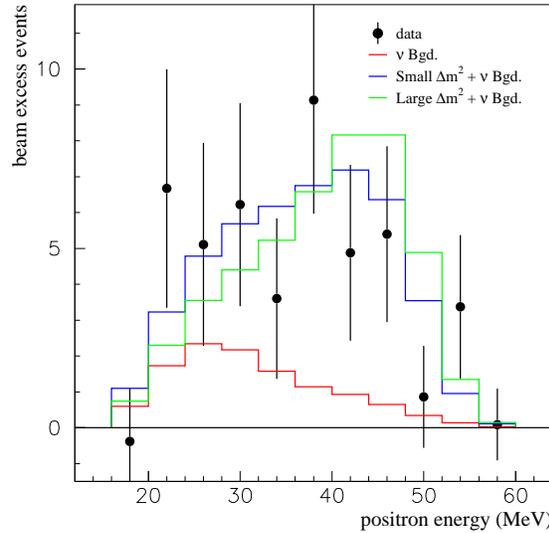,height=6.5cm}
\caption{Energy distribution of the excess $\nuebar$ interactions observed in the LSND detector}\label{fig:lsnd}
\end{center}
\end{figure} 
Neutrino oscillations in the large mixing angle region are excluded by the reactor 
experiments. The large $\Delta m^2$ region is excluded by CCFR and NOMAD 
experiments, but the possibility of the neutrino oscillations in the region
$0.3<\Delta m^2 < 2$ $\ev^2$ remains open.
\end{enumerate}

Developing a consistent interpretation of the data shown in Fig.\ref{fig:three_osc} is difficult. Three different massive neutrinos allow for only
two independent $\Delta m_{ij}$. Most of the proposed scenarios invoke
a new, hitherto unknown, sterile neutrino or postulate that some of the experimental
results are not, in fact, manifestations of the neutrino oscillations.

Given the potential importance of the neutrino mass sector, the following 
questions pose an experimental challenge in the near future:
\begin{itemize}
\item Are all three indications really examples of neutrino oscillations? \newline
Observation of  oscillatory behavior as a function of  distance and/or 
 neutrino energy, would be  particularly convincing proof. Seasonal or day-night
variation of the solar neutrino flux could serve as a proof of the solar
neutrino oscillation hypothesis. Observation of energy dependence of the disappearance of the 
$\numu$ flux is also possibility in the atmospheric or LSND regions.
\item What are the oscillation modes? \newline
The LSND result implies $\numubar \rightarrow \nuebar$ oscillations. 
What are the oscillation modes responsible for the solar neutrino
effect? The atmospheric neutrino deficit?  Are there sterile neutrinos? Preliminary
results from the SuperK experiment shown in Fig.\ref{fig:superk} disfavor
such a possibility as the dominant oscillation mode, because matter induced effects
would lead to a small reduction of the observed deficit.
\item What are the oscillation parameters? \newline
What are the patterns of the $\Delta m^2_{ij}$? 
 The mixing angles? What are the elements of the lepton mixing matrix? Are there
dominant and sub-dominant oscillation modes corresponding to each oscillation
frequency ($\Delta m^2$)?
\item The presence of the phase factor $\e^{-i\delta}$ in the mixing matrix U
implies a possibility of 
CP-violating effects in  neutrino oscillations. How large are they? What is
the value of $\delta$ ?      
\end{itemize}

Studies of oscillations in the solar neutrino region require extra-terrestrial
distances and/or very low energy neutrino sources, such as nuclear reactors. 
Oscillations in the atmospheric neutrinos region or the region
 indicated by the LSND
 experiment lend themselves to studies with  neutrino beams produced in the laboratory. Such investigations of the neutrino oscillations are an important 
part of the scientific program at Fermilab.

\section{Fermilab Accelerators and Neutrino Beams}

The flagship of the Fermilab high energy physics program is the Tevatron 
Collider. Recent upgrades of the accelerator infrastructure were specifically
designed to boost the luminosity of the Collider and at the same time to 
enable a fixed-target program, like neutrino experiments, to
be carried simultaneously with the Collider experiments. Two of the existing
Fermilab's accelerators are being used to produce neutrino beams: the 8 GeV Booster
and the newly constructed Main Injector.
\subsection{8 GeV Booster Neutrino Beam (MiniBOONE)}
The Booster is the oldest part of the of the accelerator complex at Fermilab. 
Upgraded to 400 MeV injection in 1993, it is capable of delivering up to 
$5 \times 10^{12}$ protons per pulse. In the past, it was operated at $2.5 \rm{Hz}$, but after improvements of its pulsed magnets, its repetition 
 rate can be as high as $7.5 \rm{Hz}$.

A Booster\cite{boone} neutrino beam is under construction. An 8 GeV proton beam
is extracted onto a titanium or nickel target. Two magnetic horns
 focus secondary pions and kaons, their focusing power being optimized
for 3 GeV secondaries. The secondary beam has a relatively short decay path 
which can be varied from 25 to 50 m. A variable decay path will provide 
additional information on the $\nue$ component of the beam.
  
Combining the high proton flux with
a high efficiency horn focusing will provide a  high flux neutrino beam,
yielding over 2,000,000 $\numu$ interactions per kton-year at a distance 
500 m from the source. The neutrino flux will have a maximum around 
$E_{\nu}=1~\rm{GeV}$  and an exponential high energy tail falling to the 10\% 
level at $E_{\nu}=3~\rm{GeV}$. 
 
\subsection{NuMI: Neutrino Beams from the Main Injector}

The Main Injector accelerator is a 150 GeV proton synchrotron constructed to
replace the original Fermilab Main Ring. It is expected to serve as a high
intensity, fast-cycling accelerator for antiproton production, as an
injector into the Tevatron and simultaneously to support  fixed target
experiments using 120 GeV protons. 

 It is expected that after the completion of the NuMI
construction project, the Main Injector will be able to deliver $3.6\times
10^{20}$ protons per year onto the NuMI target, in parallel with the
simultaneous support for the antiproton production for the Tevatron Collider
experiments.  

The high intensity and high repetition rate of the Main Injector offers an
opportunity for neutrino beams of unprecedented intensity, thus creating an
opportunity for long baseline neutrino oscillation experiments.  
The Main Injector will accelerate 6 batches of $8\times 10^{12}
$ protons each with a repetition rate of 1.9 secs. One of these batches will
be used for  antiproton production, while the remaining five batches will
be  extracted onto the neutrino target. 

\begin{figure}[h]
\vspace{3.3in}
\includegraphics{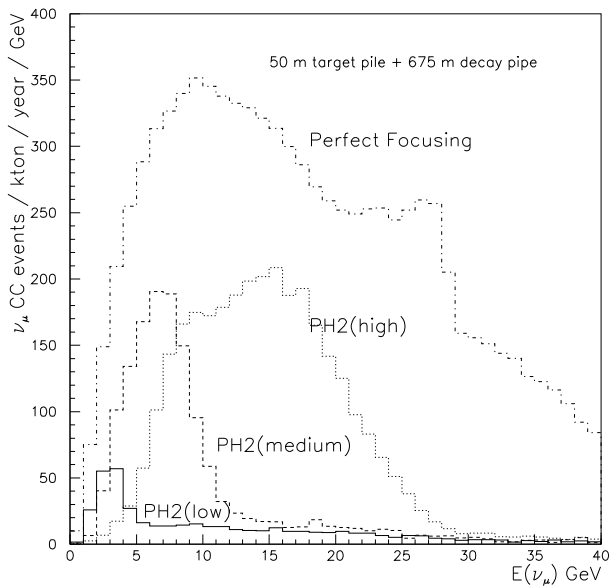}
% hsize=504 vsize=576}
\caption{Neutrino beam spectra for different NuMI beam configurations}
\label{beam}
\end{figure}

Secondary pions and kaons will be collected and focussed using a system of
two parabolic magnetic horns, and subsequently they will produce a neutrino
beam by decaying inside a 675 m long decay pipe. The beam optics is designed
to allow tuning of the neutrino beam energy by moving the focusing elements
(horns) in a manner similar to a zoom lens. The energy spectra of three
possible beam configurations are shown in Fig.\ref{beam},
 together with a
spectrum of a hypothetical beam, where all of the secondary particles were
collected and allowed to decay.  The NuMI beam design provides  overall
efficiency of the order of $50\%,$ in all three beam configurations. 

\section{MiniBOONE Experiment: Checking the LSND Result}

The primary goals of this experiment are:
\begin{itemize}
\item Unambiguously confirm or disprove the existence of the neutrino 
oscillation signal suggested by the LSND experiment
\item Provide precise measurements of the oscillation parameters, should the
existence of the effect be established, or improve the existing limits if 
the effect is not confirmed  
\end{itemize}

\begin{figure}[h]
\vspace{3.5in}
\includegraphics{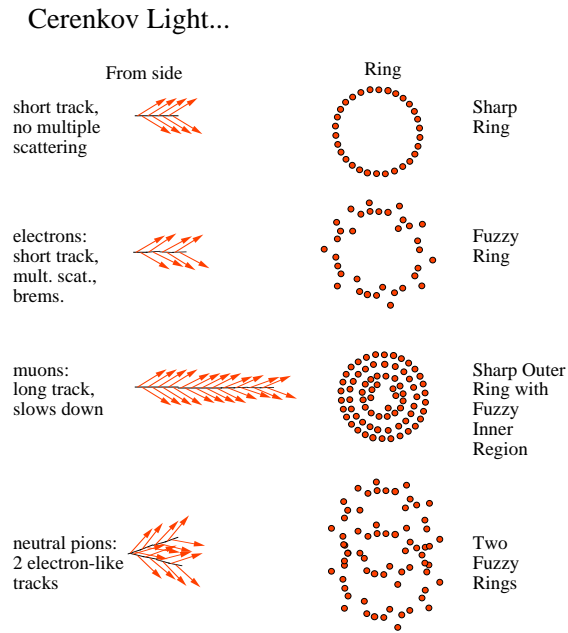}
% hsize=504 vsize=576}
\caption{Particle identification in the MiniBOONE experiement}
\label{fig:cerenkov}
\end{figure}

The detector will consist of a $12~\rm{m}$ diameter spherical tank filled with
769 tons of mineral oil,
located at the distance of $500~\rm{m}$ from the neutrino source. Cerenkov 
light emitted by  particles produced in the neutrino interactions will be
detected by 1220 eight-inch phototubes. The pattern of the detected Cerenkov rings
will be used as a primary tool of the particle ID, as shown in 
Fig.\ref{fig:cerenkov}. The outer $50~\rm{cm}$ 
volume, optically isolated from from the  main detector volume
and viewed 292 outward-pointing photomultipliers, will serve as a veto against cosmic rays.

\begin{figure}[h]
\vspace{3.5in}
\includegraphics{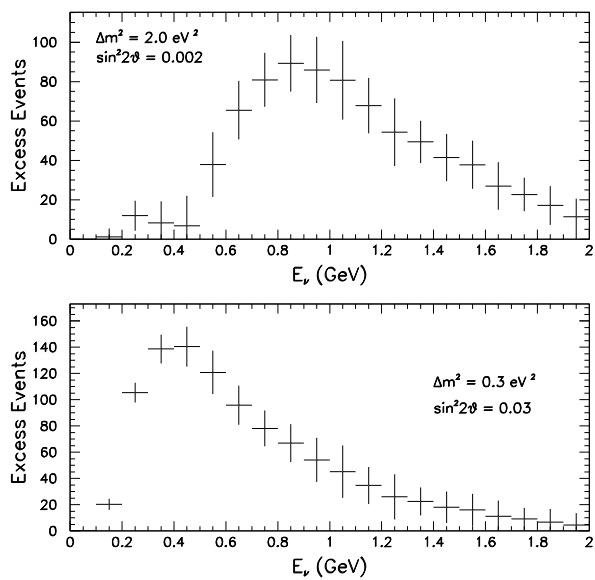}
% hsize=504 vsize=576}
\caption{Expected spectra of the excess $\nue$ interactions for two possible
oscillation scenarios}
\label{fig:en_spectra}
\end{figure}

The Booster neutrino beam is expected to yield 500,000 $\numu$ quasi-elastic 
CC interaction per year in the MiniBOONE fiducial volume. 
The expected signal of the 
neutrino oscillations, predicted
from the LSND results, will consist of a sample of 1000 identified  $\nue$ CC 
interactions. The main background will be due to  the intrinsic $\nue$ 
component of the beam: it is expected that there will be 1275 events of $\nue$
from muons decays and 425 events of $\nue$ from K decays. The background
calculations can be experimentally verified by changing the length of the decay
volume. An additional handle will be provided by the fact, that the sample
of $\nue$ interactions due  to $\numu~\rightarrow~\nue$ oscillations will have
an energy distribution different from that expected $\nue$ component of the beam (see Fig.\ref{fig:en_spectra}). 

\begin{figure}[h]
\begin{center}
\epsfig{file=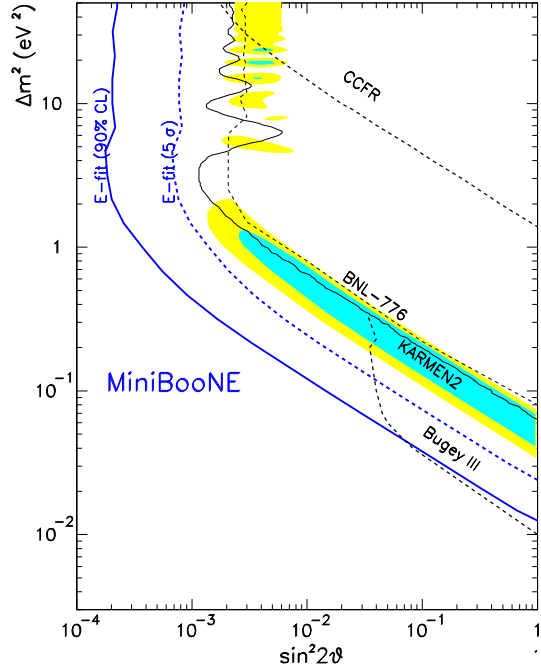,height=9.5cm}
\caption{Sensitivity of the MiniBOONE experiment}\label{fig:boone_sens}
\end{center}
\end{figure}

The large size of the expected LSND-inspired oscillation signal will enable a 
precise measurement of the underlying  oscillation parametrs whereas an 
absence of of the signal will lead to  greatly improved limits on 
possible $\numu~\rightarrow~\nue$ oscillations, shown in Fig.\ref{fig:boone_sens}.

\section{MINOS Experiment: Measuring Oscillation Parameters in the SuperK region} 
MINOS\cite{minos} experiment is designed to investigate neutrino oscillations in the
region indicated by the atmospheric neutrino experiments. Two detectors,
functionally identical, will be placed in the NuMI neutrino beam: one
at Fermilab and the second one in Soudan iron mine, 732 km away.

\subsection{ Two Detectors Neutrino Oscillation Experiment}

Two identical detectors placed in the same neutrino beam make the oscillation
experiment relatively easy. Observed interactions of $\numu$ can be divided
into two classes: ``CC''-like, with an identified $\mu$ track, and ``NC''-like,
muonless. The ratio of the observed numbers of the CC and NC-like events  in the
two detectors must be the same, provided that the same classification 
algorithm is used. This remains quite true, even if the beam spectra at the
two detector locations differ slightly, as the ratio 
$\frac{\sigma^{NC}}{\sigma^{CC}}$ is energy independent.

If $\nu _{\mu }$ undergoes oscillations, then some fraction of the original $\numu$
beam will arrive at the far detector in an 'oscillated' form. The CC 
interaction of the oscillated neutrino (with the interaction cross section
potentially reduced by a factor $\eta$ with respect to the $\sigma^{CC}_{\numu}$ ) 
will not, in general,  produce a $\mu$ in a the final state and they will  be classified 
as ``NC''-like interactions. We will have, therefore:
\begin{equation}
Near \left\{  \begin{array}{l}
NC_{near}= \Phi_{near}\left( \sigma^{NC} + \varepsilon\sigma^{CC} \right) \\
CC_{near}= \Phi_{near} \left( 1 - \varepsilon \right) \sigma^{CC}
\end{array} 
\right.
\end{equation}
and
\begin{eqnarray}
Far \left\{  \begin{array}{l}
NC_{far} = \Phi_{far}\left( \sigma^{NC} + \varepsilon\sigma^{CC} + \eta\xi\sigma^{CC} \right) \\
CC_{near}= \Phi_{far} \left( 1 - \varepsilon \right) \xi \sigma^{CC}
\end{array}
\right.
%\end{equation}
\end{eqnarray}
where: $\varepsilon$ is the fraction CC interactions misclassified as the ``NC''
events, $\xi$ is the fraction of the beam ``oscillated'' and $\Phi_{near,far}$
is the total neutrino flux at the near/far detectors.

The double ratio $R=\left(\frac{NC}{CC}\right)_{near}/\left(\frac{NC}{CC}\right)_{far}$ is 
a particularly sensitive measure of the oscillations:
\begin{equation}
R~=~\frac{\frac{1}{\xi}+\left( \varepsilon+\eta \right)\frac{\sigma^{CC}}{\sigma_{NC}}}
{1+\varepsilon\frac{\sigma^{CC}}{\sigma_{NC}}}
\label{R_def}
\end{equation}
R combines the sensitivities of the disappearance experiment, $\frac{1}{\xi}$ term, and
the appearance experiment, $\eta \frac{\sigma^{CC}}{\sigma_{NC}}$ term.
In addition R has very small systematic uncertainty, as most of the
neutrino flux uncertainties cancel. The value of R will provide  additional
information about the oscillation mode through the value of $\eta$:
\begin{equation}
\eta = \left\{ \begin{array}{ll}
1 & \mbox{$\nu_{\mu} \rightarrow \nu_e$} \\
0.2-0.3 & \mbox{$\nu_{\mu} \rightarrow \nu_{\tau}$} \\
0 &  \mbox{$\nu_\mu \rightarrow \nu_{sterile}$}
%%%%%1 & \mbox{$\nu_e}$}  
%               0.2-0.3 & \mbox{$\numu \rightarrow \nutau}$ \\
%               0 & \mbox{$\numu \rightarrow \nu_{sterile}$}
\end{array}
\right.
\end{equation}

\subsection{MINOS detectors}

The MINOS experiment will consist of two, nearly
identical detectors: one located at the Fermilab site, some 500 meters
behind the decay pipe, and the second one, in northern Minnesota, at the
distance of 732 km from Fermilab. The far detector will be located in a new
cavern, which is under construction in the Soudan mine, close to the existing Soudan
II detector.

The far MINOS detector will consist of two supermodules, 2.7 kton each. They
will be constructed as magnetized steel octagons, 8 m in diameter, with
a toroidal magnetic field about 1.5 T. Steel plates, 2.5 cm thick, will be
interspersed with planes of scintillator strips, to provide calorimetric
measurement of the deposited energy, with energy resolution $\Delta
E/E\sim 0.6/\sqrt{E}$. The active detector elements will consist of strips of
extruded scintillator, 1 cm thick and 4 cm wide. Scintillation light will be
collected by waveshifting fibers and read out by Hamamatsu M16
photomultipliers. The fine granularity of the scintillator strips will allow
them to be used as a tracking detector to measure muon trajectories and
determine the muon momentum from the curvature in the magnetic field.

The near detector, on the Fermilab site, will be as similar as possible to the
far detector, except for its size. 

The neutrino beam line and the MINOS detectors are under construction
and the data taking is expected to commence in 2003. The choice of the
initial beam energy is currently under discussion and it may depend on
the forthcoming results of the K2K experiment.

\subsection{MINOS Physics Measurements}

Two massive detectors and an intense neutrino beam constitute a powerful tool
to investigate neutrino oscillations, especially when the beam energy can
be chosen to maximize the oscillation signal. MINOS will perform several independent
measurements, which will provide a clear and complete picture of the neutrino oscillations
in the SuperK region. These measurements fall into three different categories:
\begin{itemize}
\item Firm evidence for the oscillations.

Near/far detector comparison will reduce the systematic uncertainties. 
The neutrino beam
spectrum measurement with the near detector will constrain the predicted neutrino
flux at the far detector. In the presence of the SuperK-indicated effect we expect at 
least two evidences for the oscillations:
  \begin{itemize}
    \item A double ratio  $R=\left(\frac{NC}{CC}\right)_{near}/\left(\frac{NC}{CC}\right)_{far}$
          (see Eq.\ref{R_def}) different from one. The sensitivity of this measurement for 
          different possible $\Delta m^2$ depends on the selected beam energy, as shown in 
          Fig.\ref{fig:ttest}
\begin{figure}[h]
\begin{center}
\epsfig{file=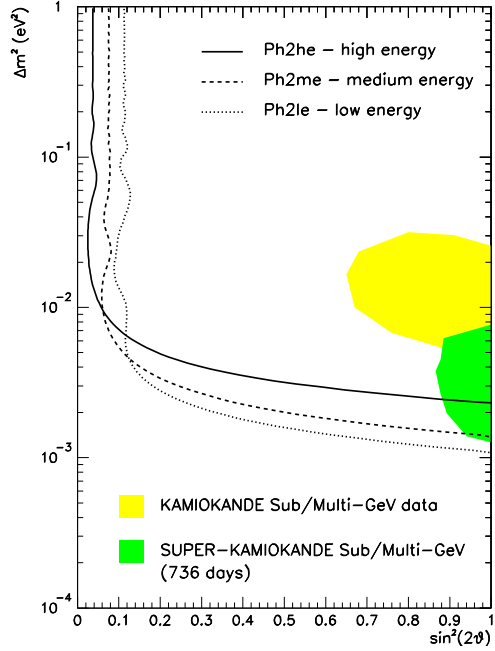,height=9cm}
\caption{90\% C.L. limits on the $\numu \rightarrow \nutau$  oscillations parameters for 2 years
exposure}\label{fig:ttest}
\end{center}
\end{figure}
     \item The $\numu$ charged curent interaction rate and the observed energy distribution.
        Presence of  neutrino oscillations will lead to a characteristic oscillatory
        modification of the spectrum observed at the far detector, as shown in Fig.\ref{fig:oscill}
\begin{figure}[h]
\begin{center}
\epsfig{file=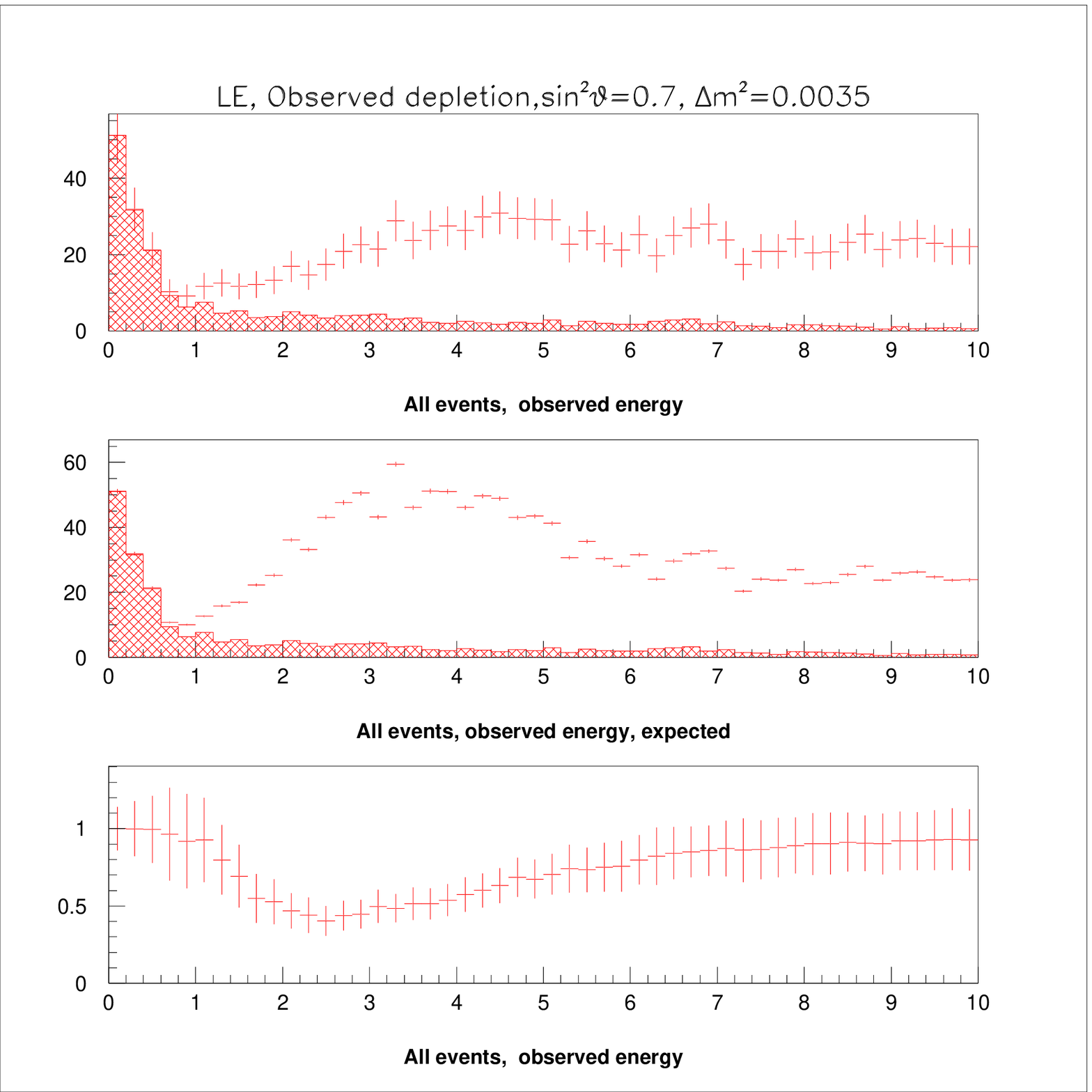,height=9cm}
\caption{Observed spectrum of the identified $\numu$ CC events in the presence (top) or
absence (middle) of oscillations. Shaded histogram represents contribution of mis-identified
NC events. Ratio of the observed and the expected distributions is shown at the bottom for
2 years exposure.}\label{fig:oscill}
\end{center}
\end{figure}
  \end{itemize}
\item Measurement of the oscillation parameters: $\Delta m^2$ and $\sin^{2}2\theta$.

Fits of the observed depletion of the CC energy spectrum in Fig.\ref{fig:oscill} will provide
a precise estimate of the oscillation parameters. The expected precision of this determination
depends somewhat on the oscillation scenario and on the choice of the beam. Two years exposure
of the MINOS detectors will yield measurements with the precision illustrated in Fig.\ref{fig:prec}.
\begin{figure}[h]
\begin{center}
\epsfig{file=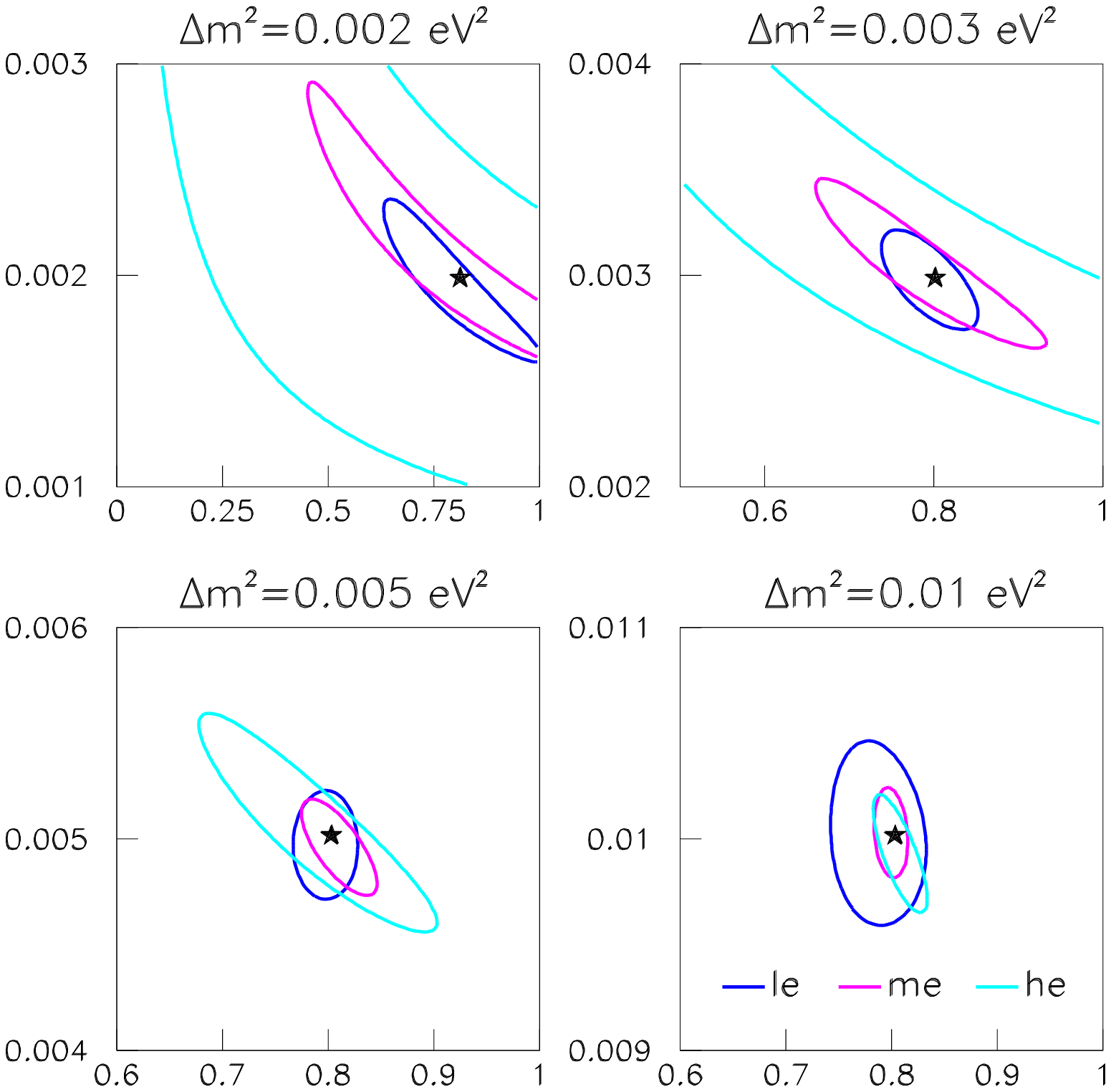,height=9cm}
\caption{The 68\% C.L. error contours for the expected oscillation signal with
$sin^2 2\theta=0.8$ and different $\Delta m^2$. Different contours represent measurements with
different beams: low (le), medium (me) and high (he) energy and two years exposure.}\label{fig:prec}
\end{center}
\end{figure}

\item Determination of the oscillation mode(s)

   \begin{itemize}
      \item $\numu \rightarrow \nu_{sterile}$ ?

	The large mixing angle indicated by the SuperK results leads to a significant contribution
        of the appearance term to R in Eq.\ref{R_def}. The measurement of R will provide a decisive 
        demonstration for or against sterile neutrinos as a dominant oscillation mode over the entire
        region of SuperK.
      \item $\numu \rightarrow \nue$ ?

	The fine granularity of the MINOS detector will allow for identification of the $\nu_e$ 
        interactions by topological criteria
         with efficiency of the order of $15-20\%$. The background of the mis-identified
        NC interactions as well as the intrinsic $\nu_e$ component of the beam (expected to be of
        the order of $0.6\%$) will be measured with high accuracy by the near detector. Fig.\ref{fig:e}
        shows the sensitivity of the MINOS detector to this oscillation mode in comparison with
        the limits from CHOOZ experiment. 
\begin{figure}[h]
\begin{center}
\epsfig{file=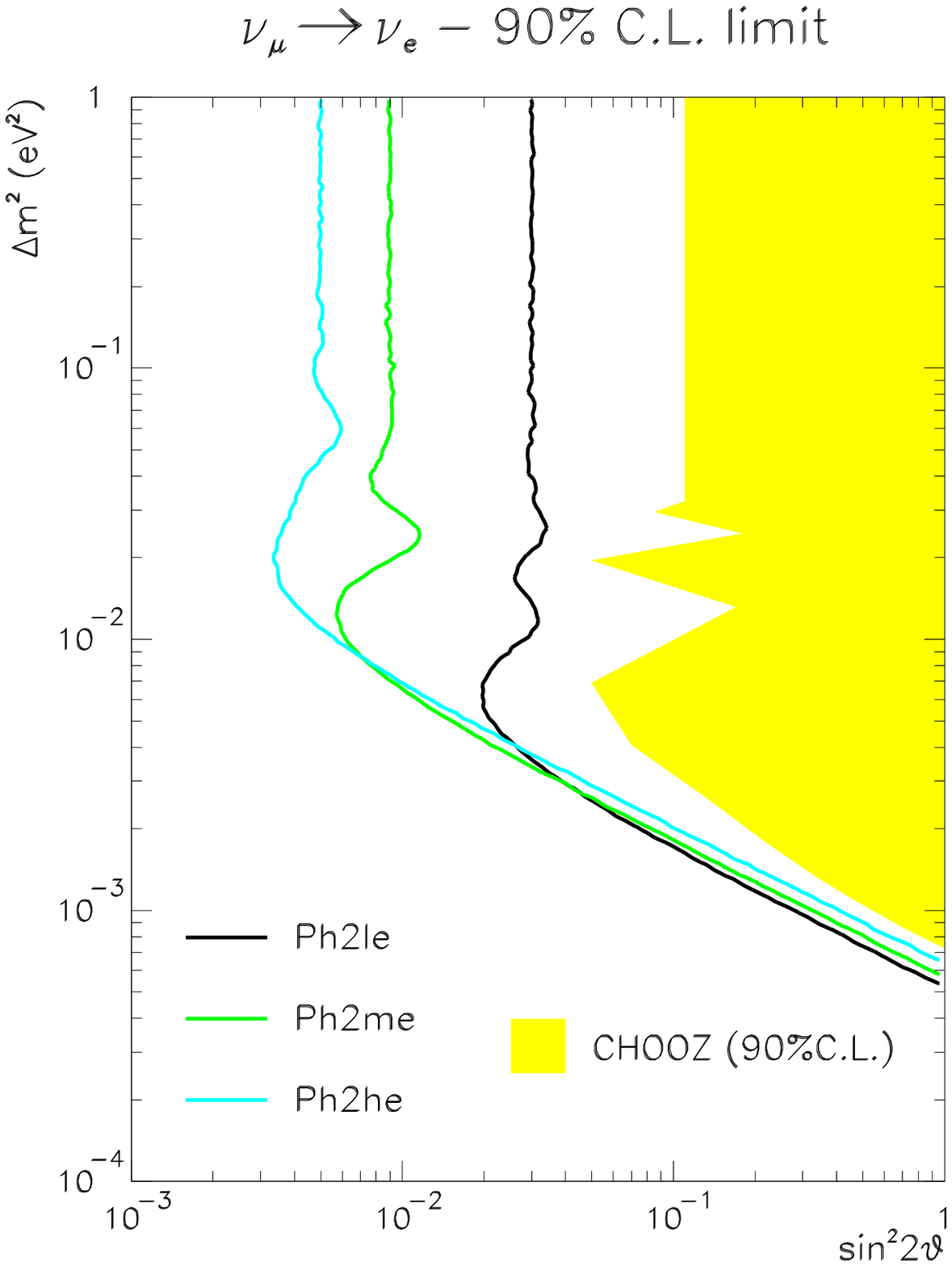,height=9cm}
\caption{The 90\% C.L. limits for $numu \rightarrow \nue$ oscillation parameters for two
years exposure with different beams: low (Ph2le), medium (Ph2me) and high (Ph2he) energy.}\label{fig:e}
\end{center}
\end{figure}

\item $\numu \rightarrow \nutau$ ?
\nopagebreak

	Circumstantial evidence for this oscillation mode will be provided by the measurement of R
(Eq.\ref{R_def}). For relatively high $\Delta m^2$, above $5\times10^{-3}~\rm{eV}^2$ a significant sample
of CC $\nutau$ interactions can be identified by exclusive decay modes, like $\tau \rightarrow \pi$.
The near detector will be again instrumental in reliable determination of the unavoidable background 
from the NC interactions.  
   \end{itemize}
\end{itemize}

\section{Conclusions and Outlook}

New experiments, under construction at Fermilab, will help to clarify the situation
with neutrino oscillations in the $\Delta m^2 > 0.001 \rm{eV}^2$ region.
The MiniBOONE experiment will settle, within coming 2-3 years, the issue of the LSND results
by precise determination of the underlying oscillation parameters or by setting limits
far outside the LSND-allowed region.
MINOS experiment  will decisively establish
the phenomenon of neutrino oscillations and measure precisely the corresponding mixing
angles and $\Delta m^2$ values in the region indicated by the SuperK experiment within 
next 5-6 years. The question 
of the dominant oscillation mode: $\numu \rightarrow \nutau$ or $\numu \rightarrow \nu_{sterile}$ will be settled. The sub-dominant mode $\numu~\rightarrow~\nue$ will be established or the existing
CHOOZ limit will be significantly improved.  The next generation of the oscillation experiments
will probably await a new generation of neutrino beams, derived from muon storage rings. 
The high intensity of such beams will make it possible to detect subtle effects like CP-vilation or matter
induced effects. By providing $\nue$ beams along with the $\numu$ component, these beams
will enable complete measurements of the neutrino mixing matrix elements.
  
\section{Acknowledgements}
It is a pleasure to thank and to congratulate
the organizers, especially Prof. M. Jezabek, for the flawless organization of such 
a pleasant and stimulating conference. Many of my collegues in MINOS collaboration
contributed to this presentation, I would like to thank in particular Dr. D. Petyt
for his contribution to understanding of the physics potential of this experiment.
Profs. J. Conrad and M. Shaevitz helped me to fully apreciate the beauty of the MiniBOONE
experiment.

Apologizes go to all those, whom I neglected to quote properly. The number of
relevant papers is overwhelming.   

%\section*{References}

\end{document}